\documentclass[conference]{IEEEtran}

\usepackage{url,comment,cite}
\usepackage{amsmath,amssymb,amsfonts,bbm}
\usepackage{graphicx}
\usepackage{epsfig,epsf,psfrag,textcomp}
\usepackage[usenames,dvipsnames]{xcolor}
\usepackage{subfigure}
\usepackage{float}
\usepackage{dblfloatfix}
\usepackage{algorithm2e,algorithmic}
\usepackage{multirow,dsfont}
\usepackage{threeparttable,tabularx}

\newcommand{\Fig}[1]{Fig.~\ref{fig:#1}}
\newcommand{\Sec}[1]{Sec.~\ref{sec:#1}}
\newcommand{\Tab}[1]{Tab.~\ref{tab:#1}}

\makeatletter
\def\bstctlcite{\@ifnextchar[{\@bstctlcite}{\@bstctlcite[@auxout]}}
\def\@bstctlcite[#1]#2{\@bsphack
  \@for\@citeb:=#2\do{%
    \edef\@citeb{\expandafter\@firstofone\@citeb}%
    \if@filesw\immediate\write\csname #1\endcsname{\string\citation{\@citeb}}\fi}%
  \@esphack}
\makeatother

\begin{document}

\bstctlcite{IEEEexample:BSTcontrol}	

\title{
What is LTE {\em actually} used for? An answer\\
through multi-operator, crowd-sourced measurement
} 

\author{
\IEEEauthorblockN{Francesco Malandrino}
\IEEEauthorblockA{The Hebrew University of Jerusalem\\
Email: francesco@cs.huji.ac.il
\vspace{-1cm}
}
\and
\IEEEauthorblockN{Scott Kirkpatrick}
\IEEEauthorblockA{The Hebrew University of Jerusalem\\
Email: kirk@cs.huji.ac.il
\vspace{-1cm}
}
\and
\IEEEauthorblockN{Danny Bickson}
\IEEEauthorblockA{The Hebrew University of Jerusalem\\
Email: danny.bickson@gmail.com
\vspace{-1cm}
}
} 

\maketitle

\begin{abstract}
LTE networks are commonplace nowadays; however, comparatively little is known about where (and why) they are deployed, and the demand they serve. We shed some light on these issues through large-scale, crowd-sourced measurement. Our data, collected by users of the WeFi app, spans multiple operators and multiple cities, allowing us to observe a wide variety of deployment patterns. Surprisingly, we find that LTE is frequently used to improve the {\em coverage} of network rather than the capacity thereof, and that no evidence shows that video traffic be a primary driver for its deployment. Our insights suggest that such factors as pre-existing networks and commercial policies have a deeper impact on deployment decisions than purely technical considerations.
\end{abstract}

\section{Introduction}
\label{sec:intro}

Cellular networks are remarkable among human-made entities in that they are both ubiquitous and inaccessible. Virtually every human being is covered by one or more cellular network; yet, it is surprisingly difficult to ascertain how exactly such networks are planned, deployed, and utilized.

In most cases, such information is simply unavailable, being a closely guarded commercial secret. While some information is available, it only concerns the deployment, i.e., the location of cellular base stations~\cite{kibilda-wd}, not the demand they serve. When demand information is available, only the aggregate traffic is known, not the applications it is made of~\cite{difra-tccn}. When detailed traffic information is available, it typically includes only one mobile operator and limited geographical scope~\cite{urbane-orange}.

The heart of the matter is that mobile network operators are not necessarily the best-suited observers to collect information on mobile networks. A powerful alternative is represented by {\em crowd-sourced} network measurement, where volunteers run an application on their mobile devices, and the logs generated by such applications are combined together. Crowd-sourced measurement have existed for decades~\cite{mit-reality,infocom06-trace}; however, only recent projects based on smartphones and apps have been able to attain {\em both} a high level of detail and large scale.

There are three main aspects of crowd-sources traces that cannot be matched by traces collected by mobile operators. First and most obviously, crowd-sourced traces naturally include information about multiple operators, collected at the same time and location and with the same methodology. Furthermore, they include detailed information about the traffic demand and the individual app generating it, which mobile operators cannot obtain or share. Finally, they reveal what people do when they are not using a cellular network, including detailed mobility (as opposed to coarse, cell-level one) and information about Wi-Fi traffic. On the negative side, they are only collected by users of a certain application: the bias of such a self-selected sample and its size can be problems.

Given that crowd-sourced traces contain more information than other ones, it is natural to wonder how such information can be used. One use of our traces that is especially relevant to researchers is checking (and correcting, if need be) the assumptions made in studies on next-generation network planning. The problem is hugely relevant and widely popular; however, most studies could benefit from a better understanding of how -- and for which purpose -- present-day networks are designed and deployed.

Our paper is organized as follows. After reviewing related work in \Sec{relwork}, in \Sec{dataset} we introduce our dataset, highlighting how it compares to similar traces existing in the literature. \Sec{process} recaps the information we need to extract from our data, and summarizes the tools and methodologies we employ to that end. In \Sec{findings} we summarize our main findings, comparing them with widespread assumptions and common (mis)conceptions. \Sec{conclusion} concludes our paper, describing our ongoing efforts to make at least some of our information available to the community.

\section{Related work}
\label{sec:relwork}

\begin{figure*}[b!]
\centering
\begin{minipage}{.27\textwidth}
\includegraphics[width=.81\textwidth]{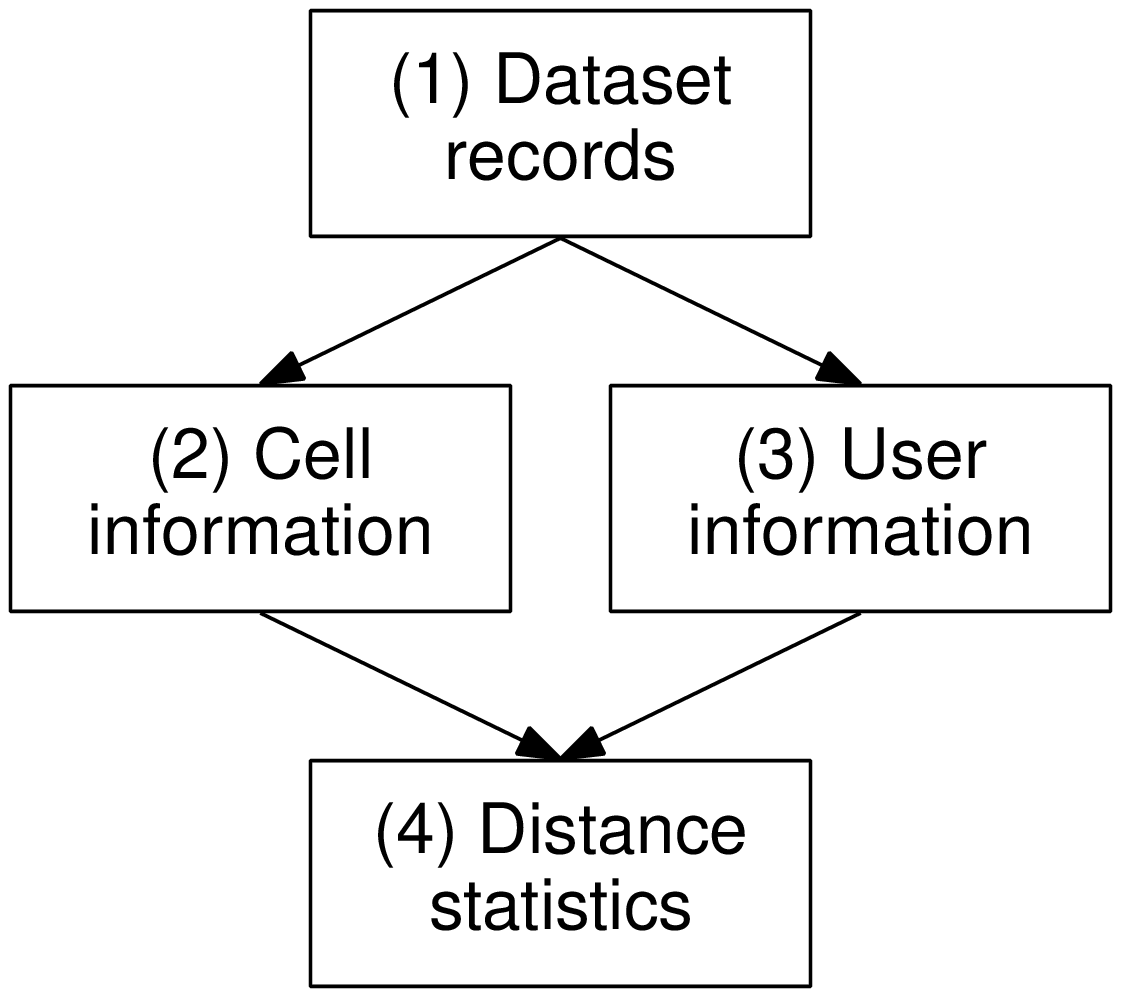}
\caption{Processing steps.
\label{fig:steps}
} 
\end{minipage}
\hspace{1em}
\begin{minipage}{.63\textwidth}
\subfigure[\label{fig:cell-traffic}]{
\includegraphics[width=.5\textwidth]{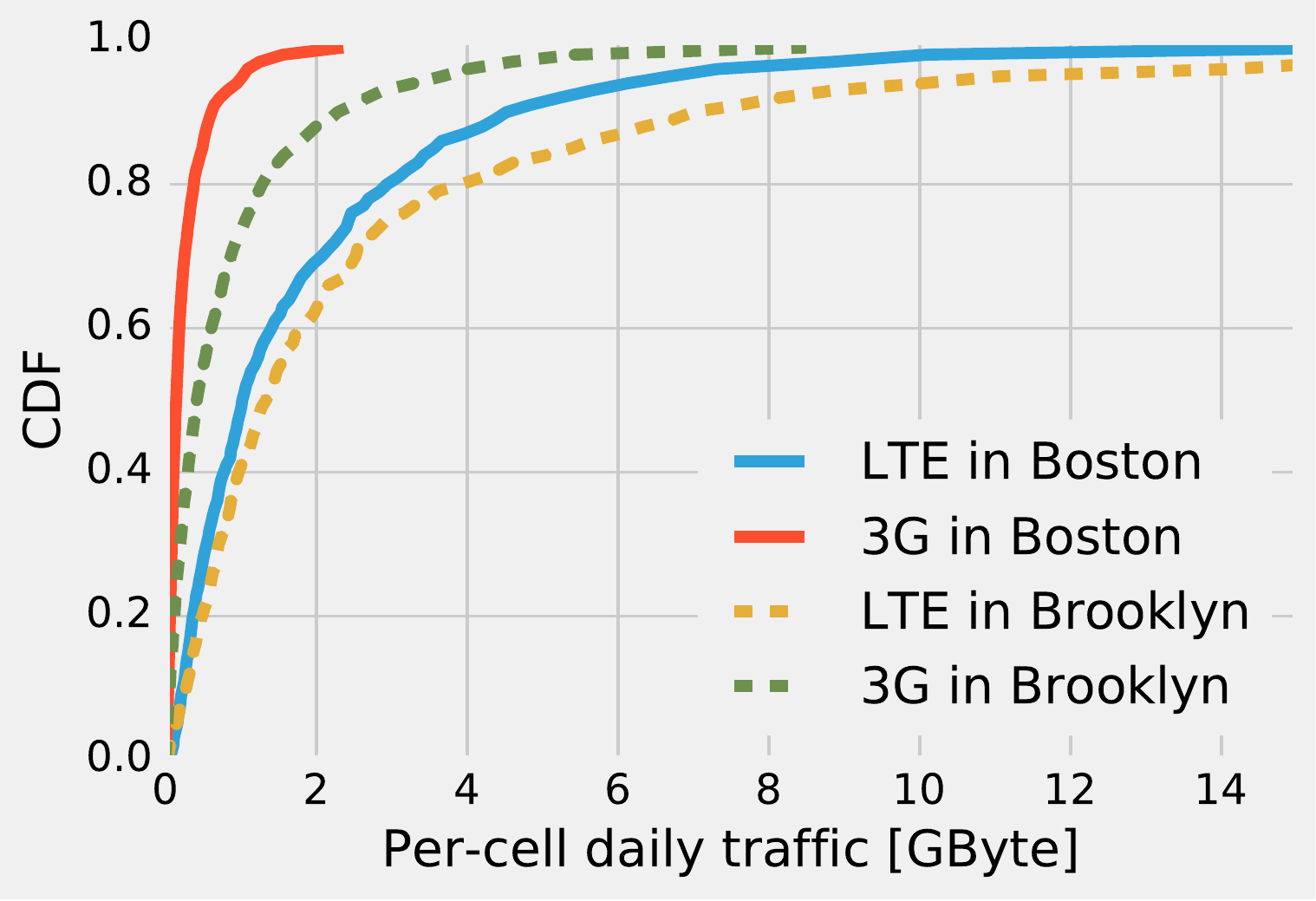}
} 
\subfigure[\label{fig:user-traffic}]{
\includegraphics[width=.5\textwidth]{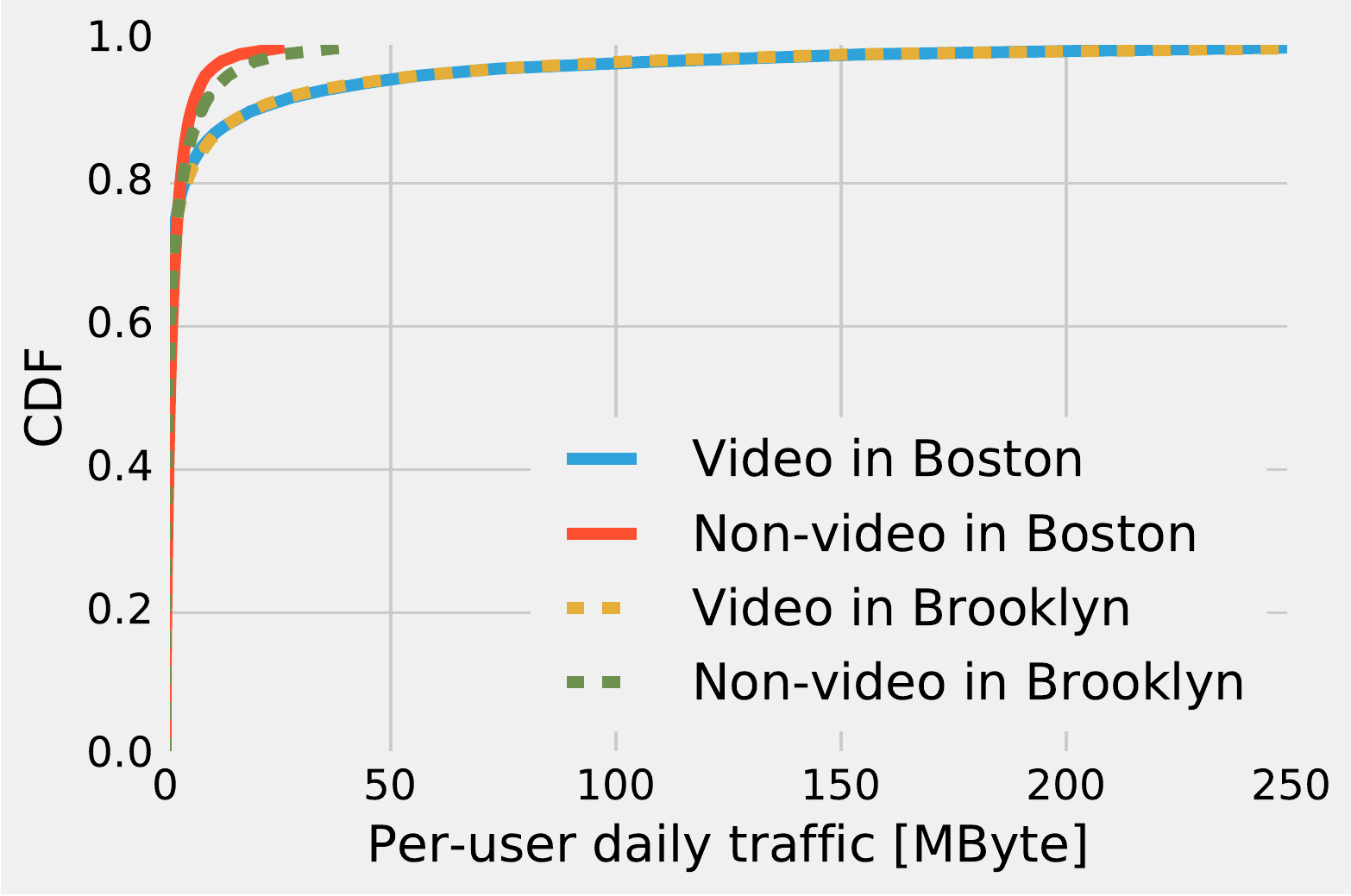}
} 
\caption{Per-cell (a) and per-app (b) traffic in the Boston and Brooklyn datasets.
\label{fig:traffic}
}
\end{minipage}
\end{figure*}

\begin{figure*}[t!]
\centering
\subfigure[\label{fig:n-cells}]{
\includegraphics[width=.3\textwidth]{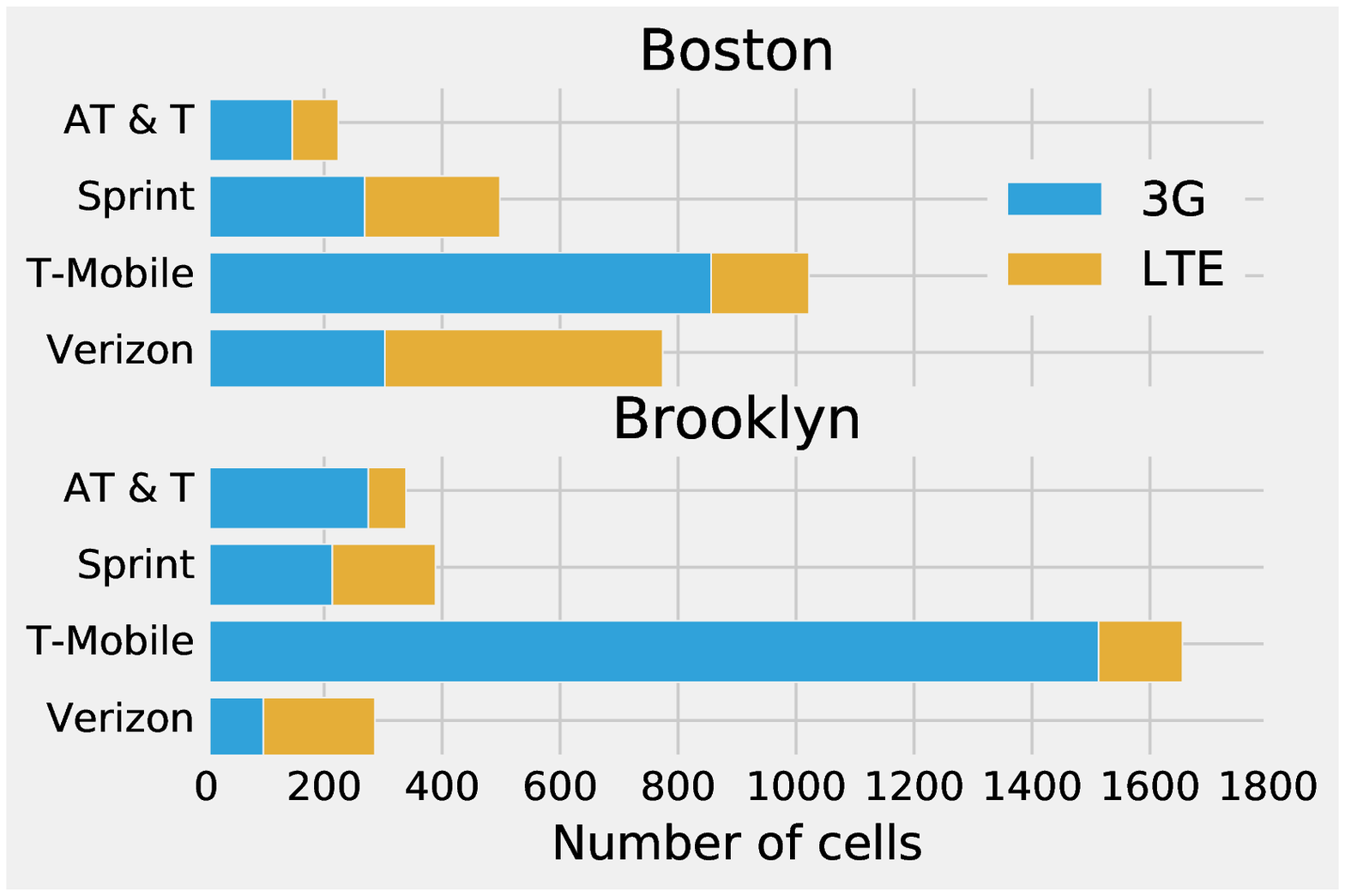}
} 
\subfigure[\label{fig:cov-boston}]{
\includegraphics[width=.3\textwidth]{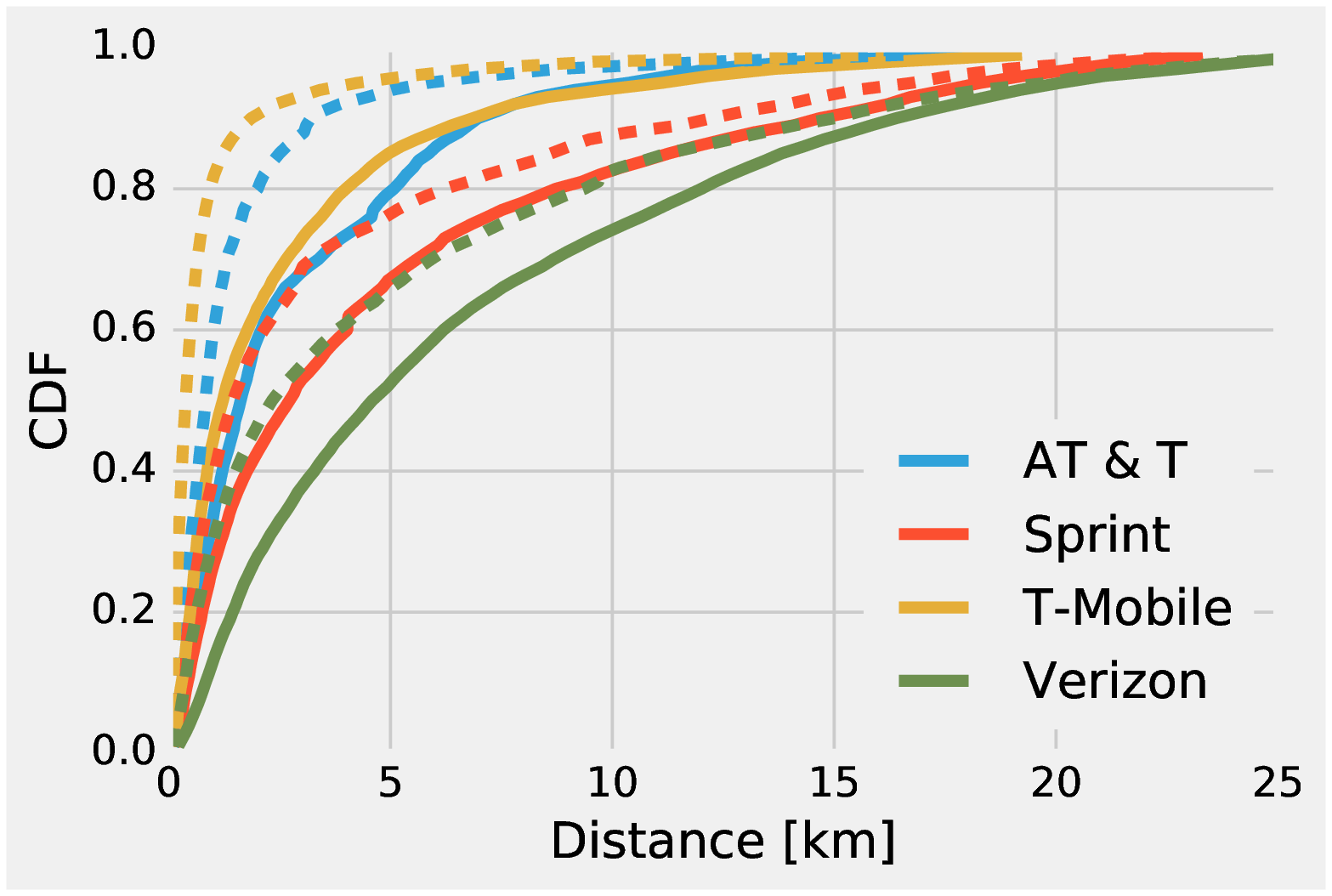}
} 
\subfigure[\label{fig:cov-brooklyn}]{
\includegraphics[width=.3\textwidth]{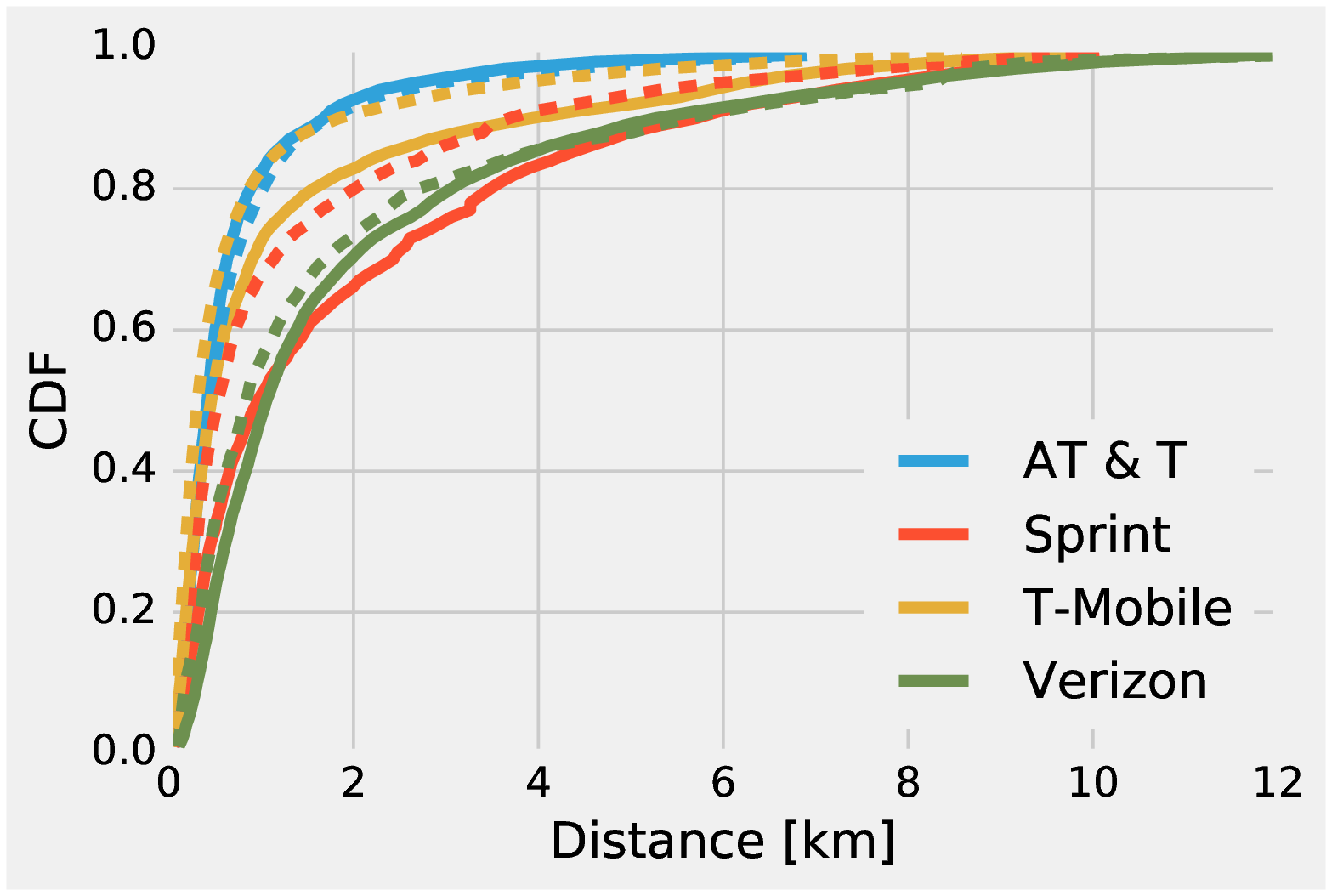}
} 
\caption{Number of LTE and 3G cells deployed by each operator (a); distance between each user and the cell covering her in Boston (b) and Brooklyn (c). Solid lines refer to LTE, dashed lines to 3G.
\label{fig:coverage}
}
\end{figure*}

\begin{figure*}[b!]
\centering
\subfigure[\label{fig:map-brooklyn-att}]{
\includegraphics[width=.23\textwidth]{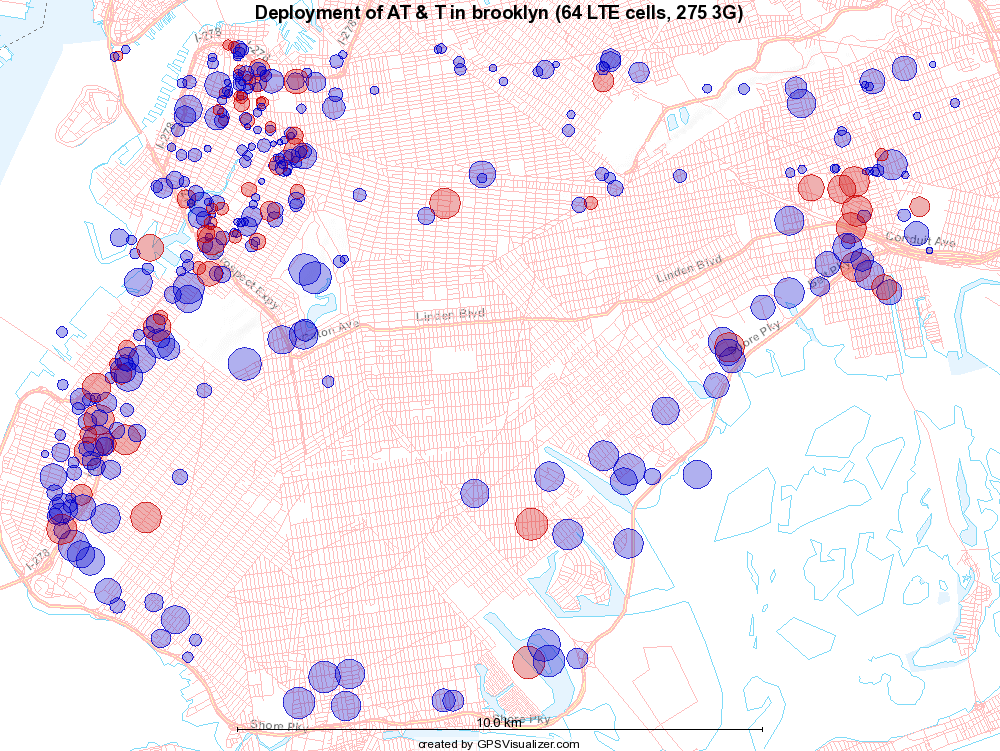}
} 
\subfigure[\label{fig:map-brooklyn-sprint}]{
\includegraphics[width=.23\textwidth]{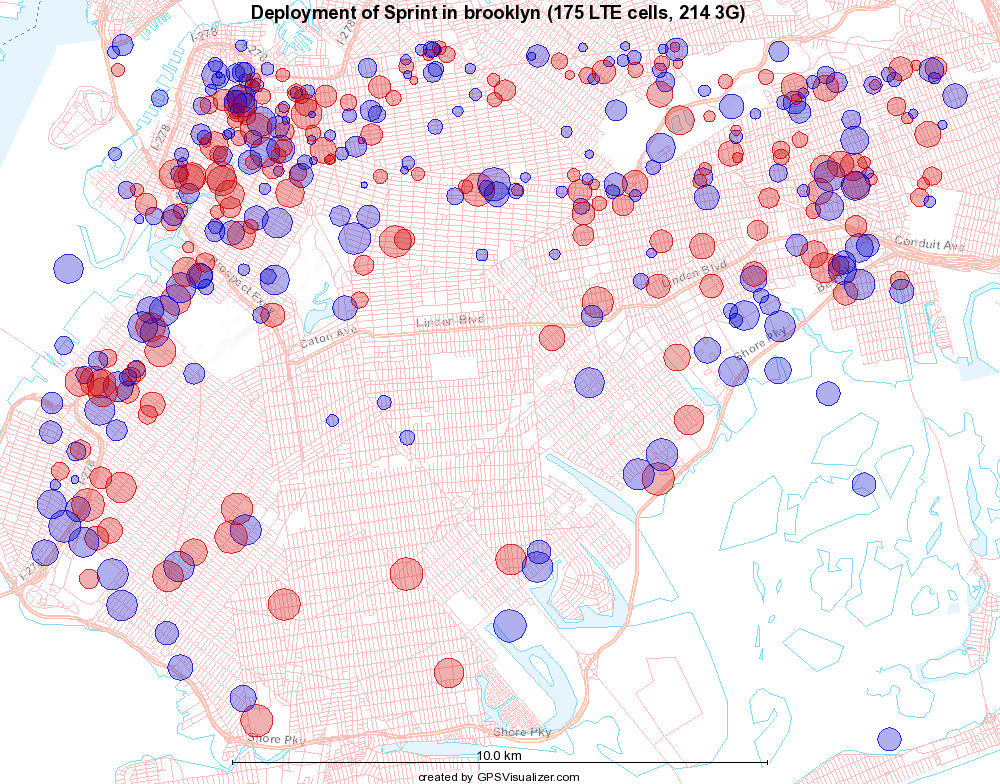}
} 
\subfigure[\label{fig:map-brooklyn-tmobile}]{
\includegraphics[width=.23\textwidth]{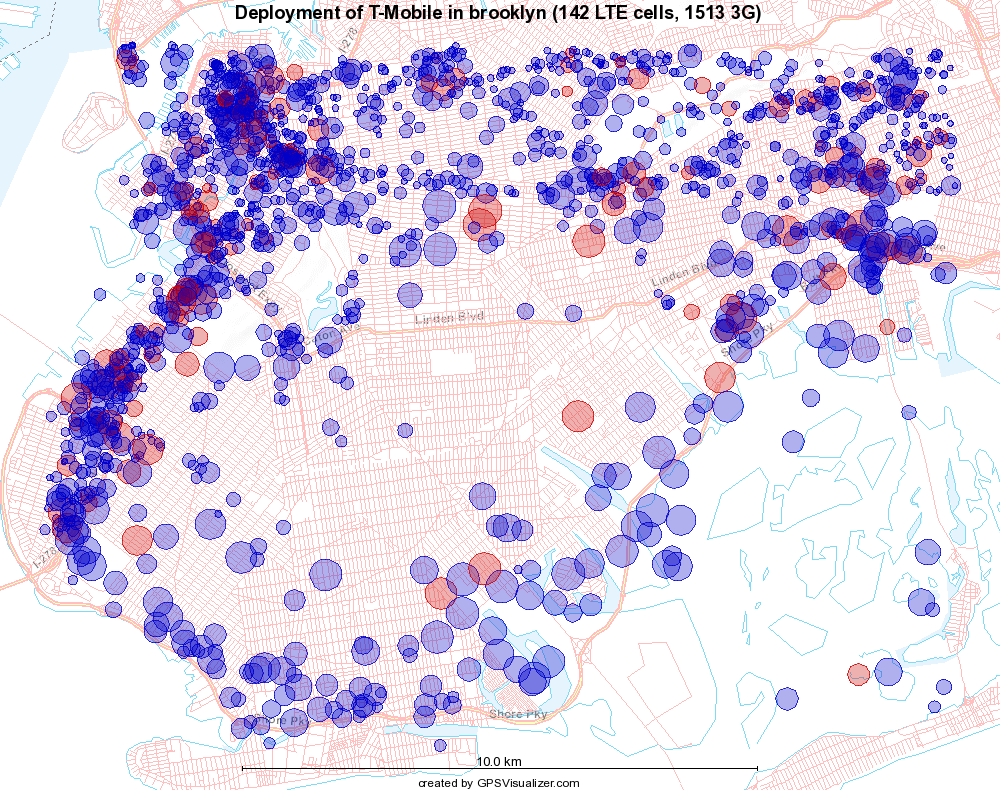}
} 
\subfigure[\label{fig:map-brooklyn-verizon}]{
\includegraphics[width=.23\textwidth]{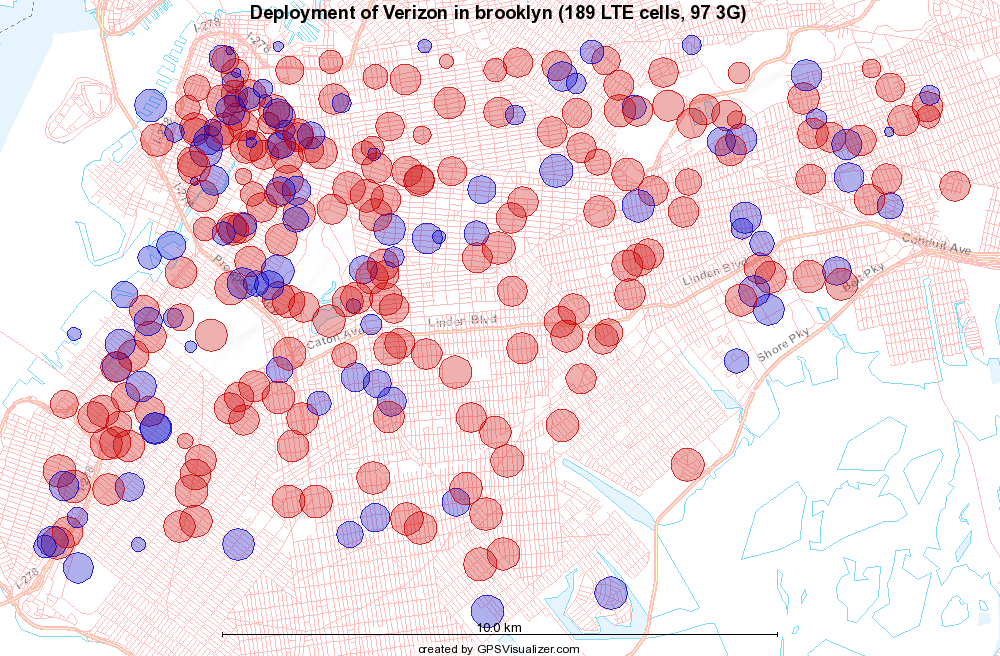}
} 
\caption{Brooklyn: deployment of 3G (blue dots) and LTE (red dots) for AT\&T (a), Sprint (b), T-Mobile (c), Verizon (d). The size of dots is proportional to the coverage area of each cell.
\label{fig:maps-brooklyn}
}
\end{figure*}

\begin{figure*}[t!]
\centering
\subfigure[\label{fig:map-boston-att}]{
\includegraphics[width=.23\textwidth]{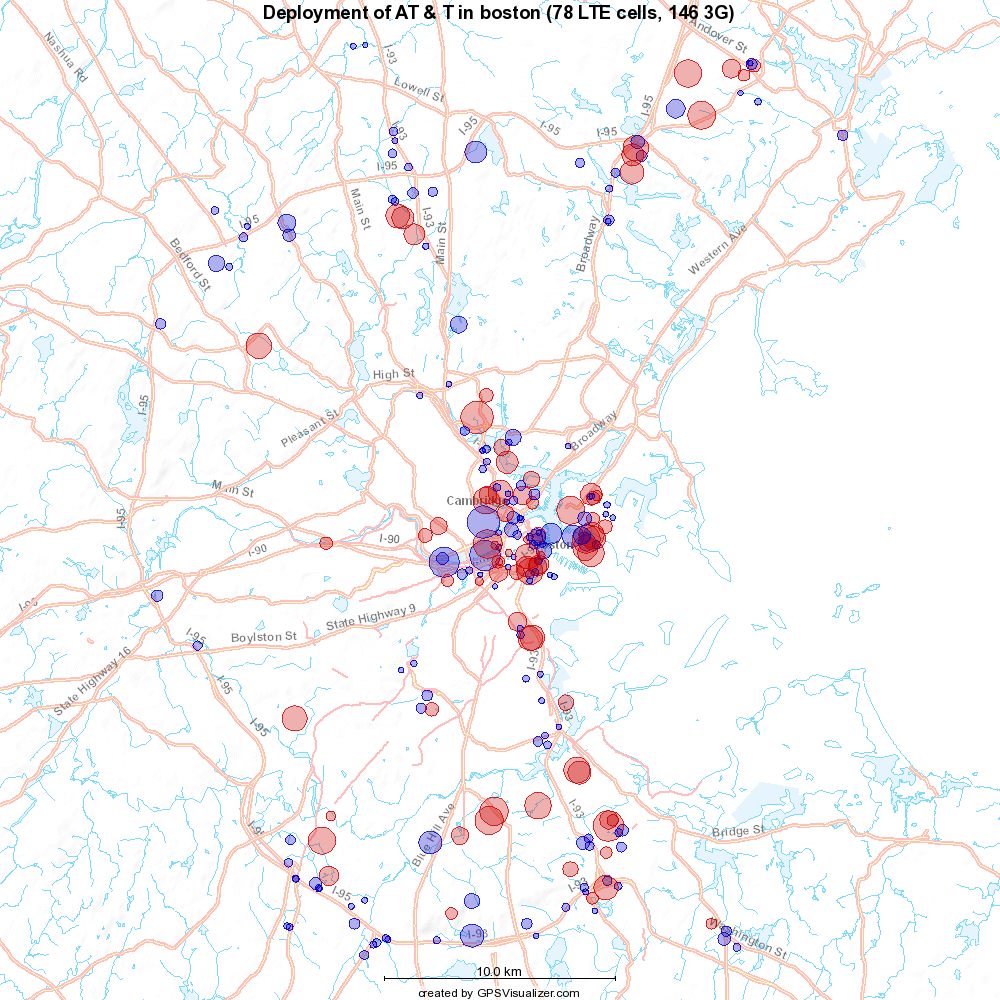}
} 
\subfigure[\label{fig:map-boston-sprint}]{
\includegraphics[width=.23\textwidth]{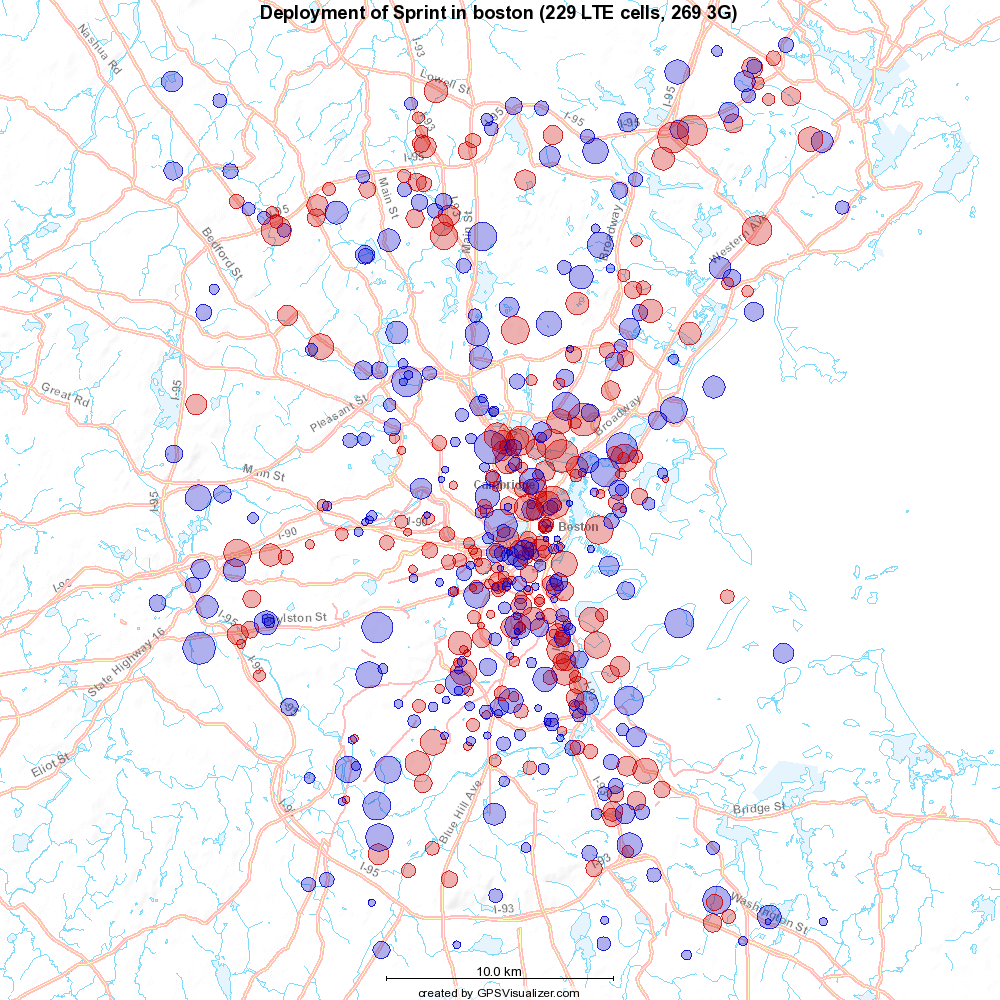}
} 
\subfigure[\label{fig:map-boston-tmobile}]{
\includegraphics[width=.23\textwidth]{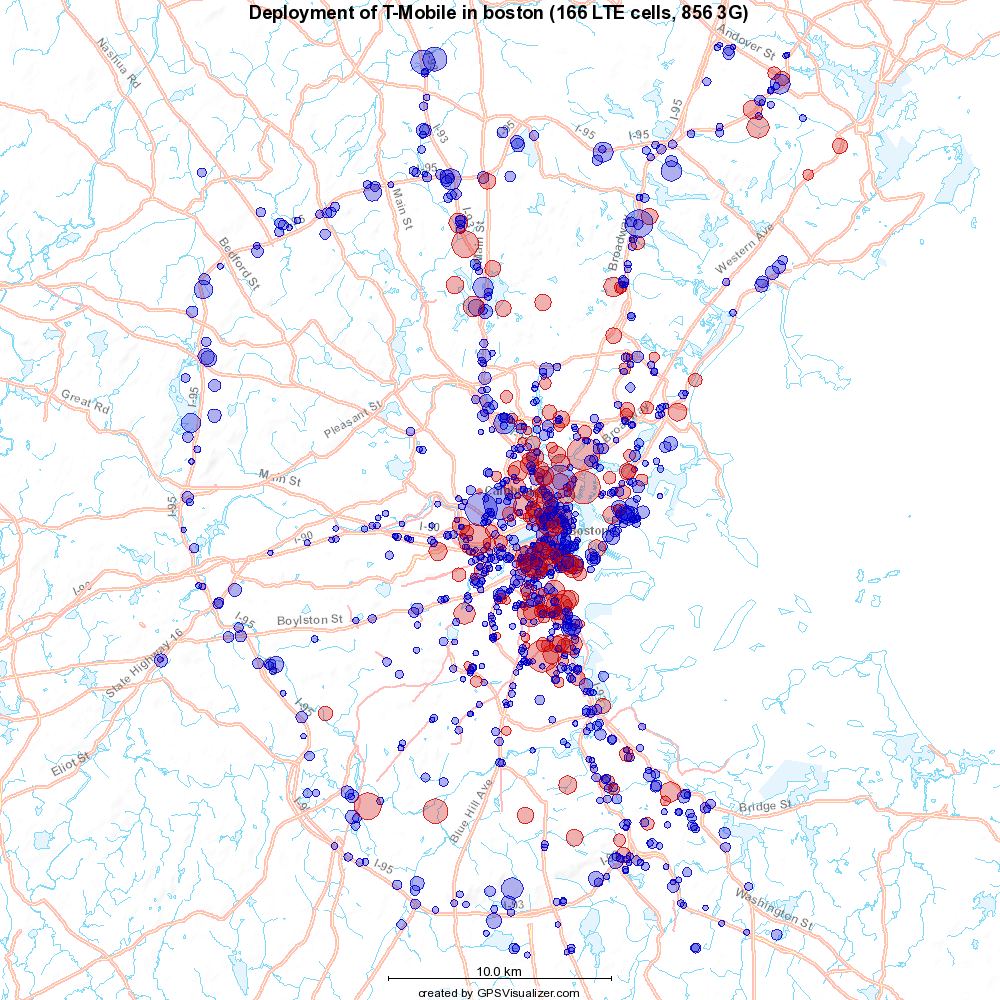}
} 
\subfigure[\label{fig:map-boston-verizon}]{
\includegraphics[width=.23\textwidth]{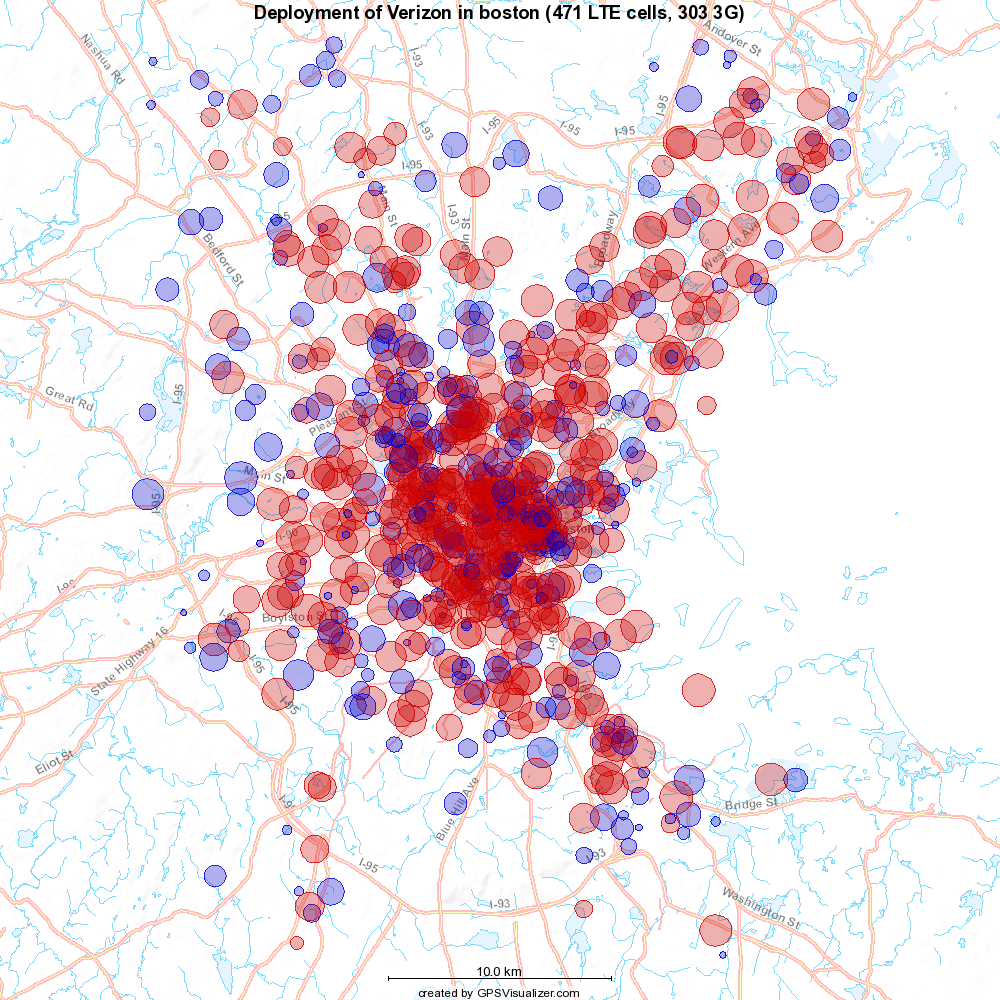}
} 
\caption{Boston: deployment of 3G (blue dots) and LTE (red dots) for AT\&T (a), Sprint (b), T-Mobile (c), Verizon (d). The size of dots is proportional to the coverage area of each cell.
\label{fig:maps-boston}
}
\end{figure*}

\begin{figure*}[b!]
\centering
\subfigure[\label{fig:serve-boston}]{
\includegraphics[width=.3\textwidth]{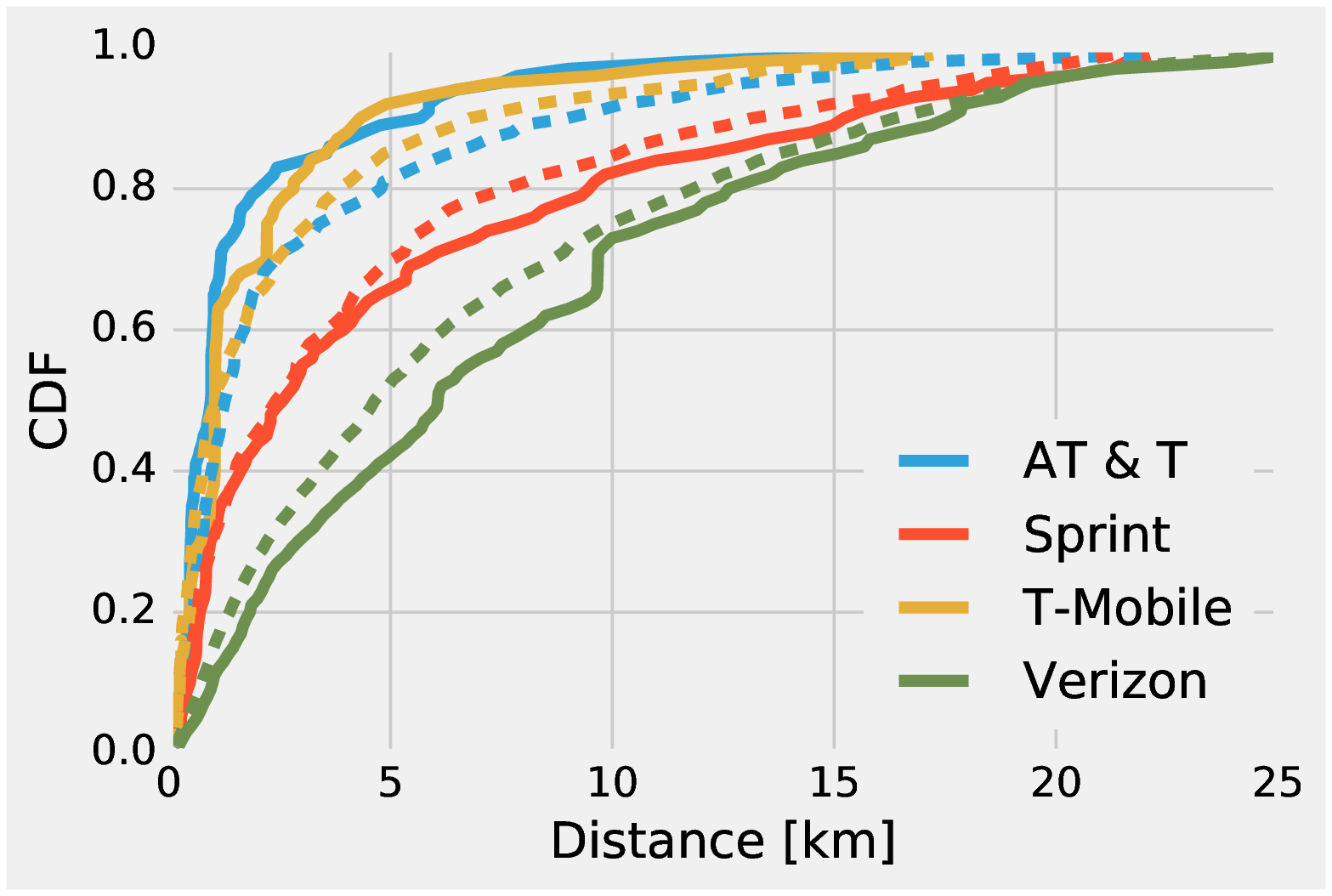}
} 
\subfigure[\label{fig:serve-brooklyn}]{
\includegraphics[width=.3\textwidth]{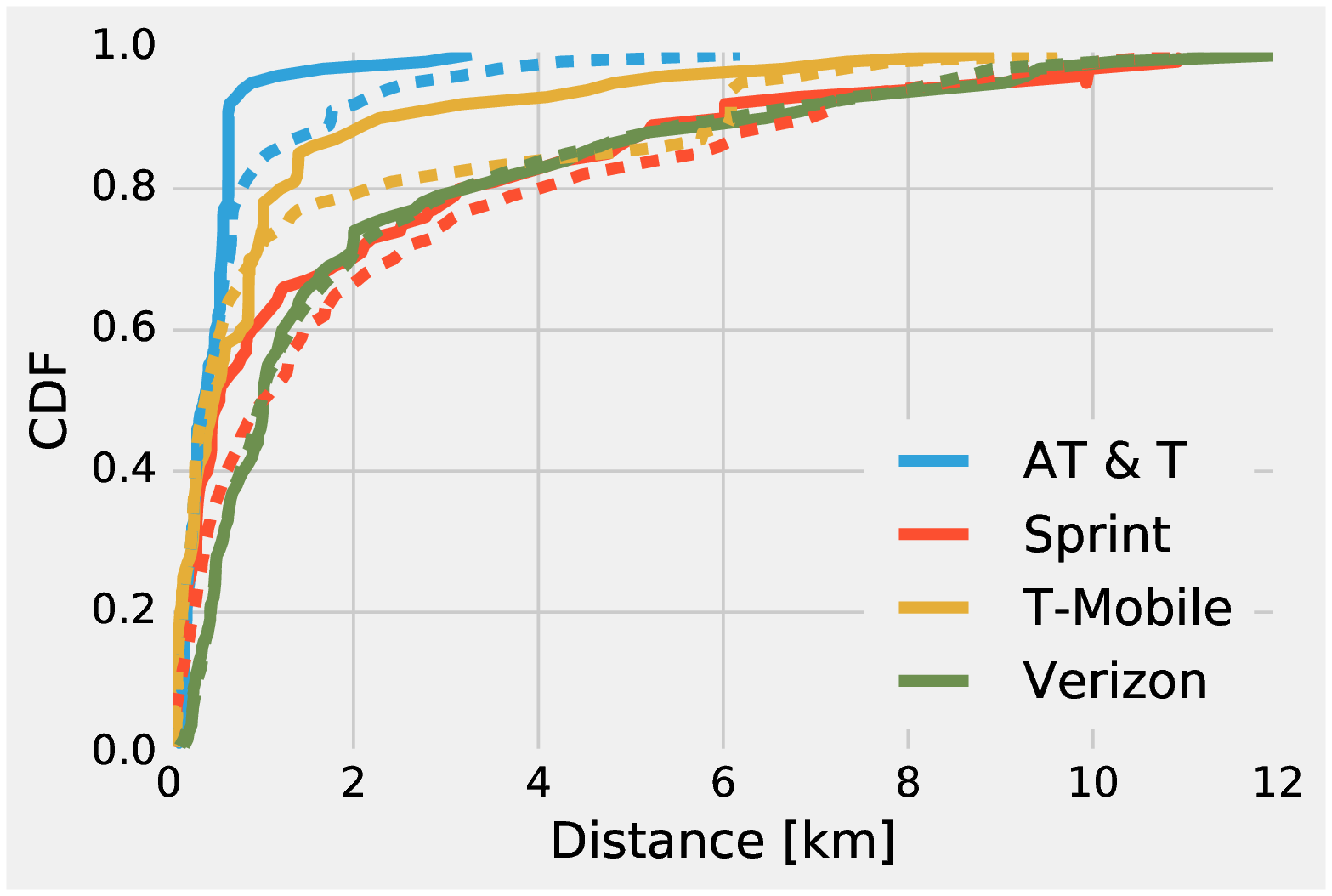}
} 
\subfigure[\label{fig:video}]{
\includegraphics[width=.3\textwidth]{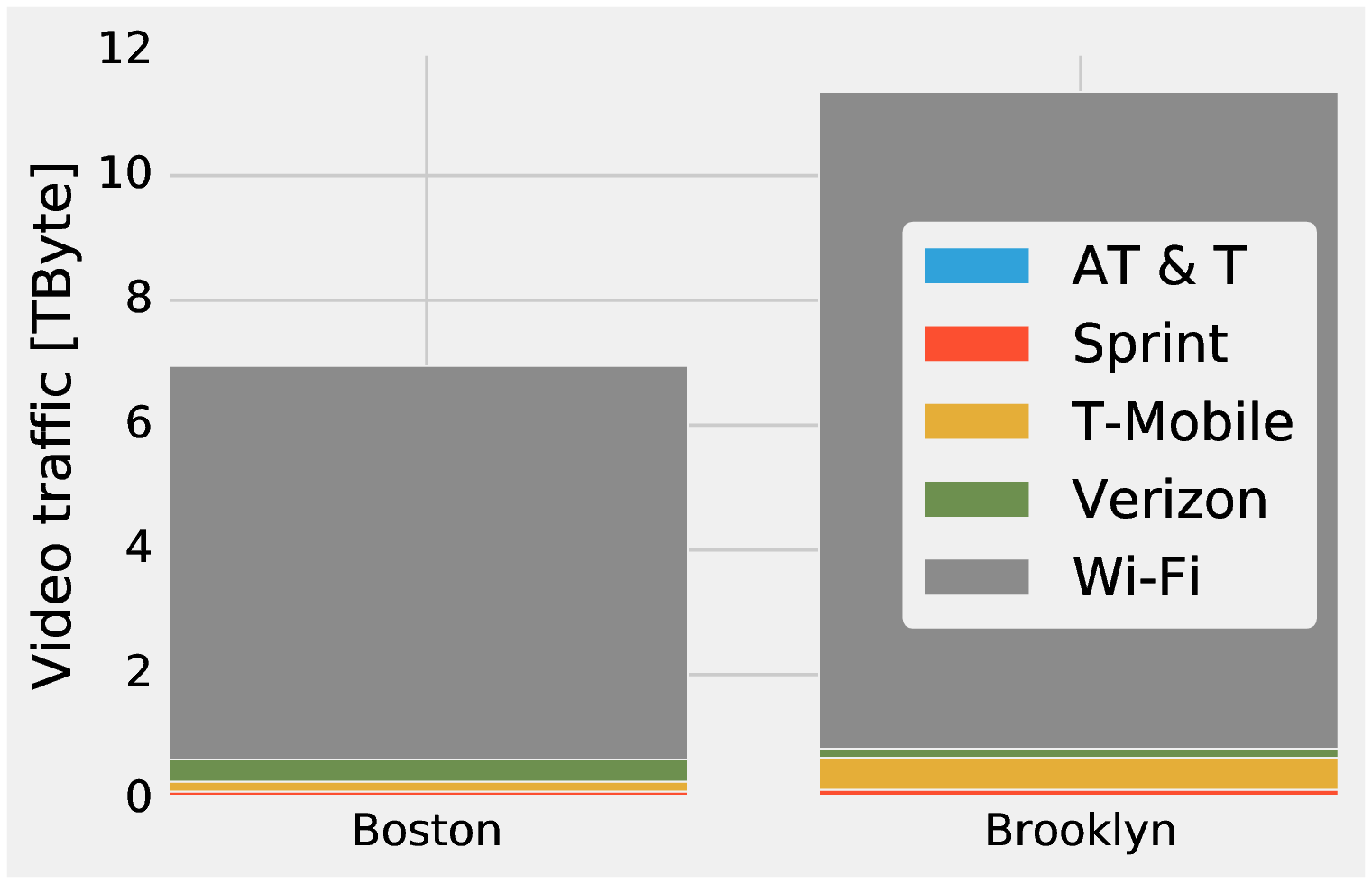}
} 
\caption{Distance traveled by LTE video and non-video traffic in Boston (a) and Brooklyn (b); operators serving video traffic (c). Solid lines refer to video traffic, dashed lines to non-video traffic.
\label{fig:distance}
}
\end{figure*}

Our results connect to three main categories of prior work: works presenting real-world mobile traces and datasets; works studying the deployment of cellular networks; and works doing the latter using the first.

Many real-world traces come from volunteers, e.g., the MIT Reality Project~\cite{mit-reality} and the Nokia Mobile Challenge~\cite{nokia}. These traces include a great deal of valuable information; however, their main shortcoming is the limited number of participants (in the case of the Nokia Mobile Challenge, around two hundred). This scale is adequate to study, for example, user mobility or encounter patterns; however, studying a whole cellular network requires information about many more users.
Mobile operators are typically reluctant to release demand and deployment information to the scientific community. An exception is represented by the Data For Development dataset by Orange~\cite{orange-d4d}, including mobility information for 50,000 users in Ivory Coast, as well as CDR (call-detail record) information for phone calls and SMS messages. This trace lacks data sessions, and is severely restricted by heavy anonymization — each ID encountered gets a new coded identity for each “ego site” to which they are a neighbor.  This makes it impossible to trace out the entire social network.

Cellular network planning is a topic of great theoretical interest and practical relevance. Existing works go to great lengths to optimize the power control~\cite{planning-capone}, and noise level~\cite{planning-abdel} of cellular networks, accounting for both the density of the users~\cite{planning-lte} and their trajectories~\cite{planning-infocom15}. Because of the limited availability of real-world information about cellular networks, however, these works cannot account for the fact that deployment decisions are based on more factors than the users and their traffic demand.

Some works~\cite{kibilda-wd,difra-tccn} do study network planning using real-world information. In particular, the authors of~\cite{kibilda-wd} find that the coverage of different cellular networks is highly redundant and significant gains could be obtained through consolidation. \cite{difra-tccn}~studies the deployment of LTE accounting for the existing 3G networks, the locations of users, and their demand. The main limit of these works is that no distinction could be made between different types of users and demand.

\section{Our data}
\label{sec:dataset}

Our data comes from the users of an app called WeFi~\cite{wefi}. The WeFi app provides its users with information on the safest and fastest Wi-Fi access points available at the user's location. At the same time (and with their consent), it collects information about the user's location, connectivity and activity.

Wefi reports over seven million downloads of the app throughout all the world, and over three billion daily records. We use two datasets, relative to the American city of Boston and {\em borough} of Brooklyn. Their main features are summarized in \Tab{datasets}.

\begin{table}[t!]
\caption{
The Brooklyn and Boston datasets.
\label{tab:datasets}
} 
\begin{tabularx}{1\columnwidth}{|c|X|X|}
\hline
& Brooklyn & Boston \\
\hline\hline
Time of collection & Nov. 2014 & Nov. 2014\\
\hline
Total traffic [TB] & 39 & 23.2\\
\hline
Number of records & 59 million & 69 million\\
\hline
Unique users & 19315 & 15629\\
\hline
Estimated coverage & 2\% & 4\%\\
\hline
Unique cells & 25971 & 20754\\
\hline
\end{tabularx}
\end{table}

Each record contains the following information:
\begin{itemize}
\item day, hour (a coarse-grained timestamp);
\item anonymized user identifier and GPS position;
\item network operator, cell ID, cell technology and local area (LAC) the user is connected to (if any);
\item Wi-Fi network (SSID) and access point (BSSID) the user is connected to (if any);
\item active app and amount of downloaded/uploaded data.
\end{itemize}
If the location of the user or the networks she is connected to change within each one-hour period, multiple records are generated. Similarly, one record is generated for each app that is active during the same period.

Similar to other crowd-sourced traces, our datasets cover multiple mobile operators and multiple technologies, including Wi-Fi. Since we have detailed information about the actual cell each user is connected to, and the technology (e.g., LTE or UMTS or CDMA) thereof, this makes our dataset especially useful to study the {\em deployment} of cellular networks of different generations. Finally, thanks to its coverage, we can observe a representative fraction of the network traffic, and virtually all network infrastructure.

\section{Processing tools and steps}
\label{sec:process}

In this section, we describe the software tools we use to process our datasets (namely, the Graphlab library and the SFrame object) and the actual processing steps we perform.

\subsection{Processing tools: SFrame}

At tens of millions of rows each, our datasets definitely qualify as ``big data'', impossible to load into the memory of any workstation and even most servers. However, this does not necessarily mean that we need a cluster of (virtual) machines to process our data; indeed, we can get our job done with a single computer and {\em scalable} computing.

Specifically, we resort to a Python library called Graphlab Create~\cite{graphlab}, offering a scalable data type called {\sf SFrame}, whose interface is similar to R's {\sf data.frame} type or Pandas' {\sf DataFrame}. The relevant difference is that SFrames are stored on disk, with portions thereof loaded into memory only when needed, i.e., when performing some operations -- thus preventing memory from limiting the amount of data being processed. Thanks to the scalability of SFrames, we are able to perform all our computations on a single computer -- albeit a powerful one, with 64 cores and 128 GB of memory.

Graphlab is a commercial product, offering a free academic license. Furthermore, all the SFrame functionality we use in this paper is available as an open-source project~\cite{sframe-github}.

\subsection{Processing steps: deployment and demand}

There are two main aspects of the networks we are interested in studying: demand, i.e., the traffic users want to consume, and deployment, i.e., the base stations they use to this end. Our starting point is represented by the raw records in the datasets we described in \Sec{dataset} (block 1 in \Fig{steps}).

The information we need about cells (block 2 in \Fig{steps}):
\begin{itemize}
\item its location and coverage area;
\item its technology and served traffic.
\end{itemize}
The latter can be simply observed from the records, which include information about the amount of transferred data and cell technology. We observe that to each cell identifier corresponds exactly one technology; it is indeed common to have multiple cells with different technologies on the same tower, but in this case they have different identifiers. \Fig{cell-traffic} shows the CDF of the amount of traffic served by cells; notice how LTE cells seem to be more loaded than 3G\footnote{
Throughout this paper, we will designate as ``3G'' both UMTS/HSDPA and CDMA technologies.} ones, and cells in Brooklyn more than those in Boston.

Reconstructing information about the location and coverage area of cells requires some additional care. From the records, we know the users’ positions when being covered (i.e., registered with) and/or served (i.e., exchanging data with) a given cell. All these locations belong to the coverage area of the cell. We go a step further, and assume that the {\em convex hull} of all such locations corresponds to the cell's coverage area, and the base station itself is located at the baricenter of the hull. Both assumptions imply some loss of precision; however, they allow us to classify the cells according to the area they cover, and to study the distance between the base stations and the users they serve, as we will see in \Sec{findings}.

The main aspect of user traffic (block 3 in \Fig{steps}) we are interested in is its type. In particular, because video is widely expected to make up most of the demand of future networks, and indeed considered one of the main reasons why we need future networks at all, we distinguish the user demand in ``video'' and ``non-video''. We obtain this information from the active app at every record: apps such as YouTube, Hulu and Netflix all belong to the first category. \Fig{user-traffic} shows that, indeed, video apps require much more data than others.

At a first sight, \Fig{traffic} seems to confirm our intuitive expectations. Video applications require large amounts of data, and new, high-speed LTE base stations carry most of it, concerning the {\em matching} between the traffic demand shown in \Fig{user-traffic} and the deployment summarized in \Fig{cell-traffic}. Are LTE networks deployed to improve the ability of cellular networks to serve video traffic? Are they better-suited to video traffic than to other types of data? Do different mobile operators have different strategies in this respect?

To answer these questions we need to move to block 4 in \Fig{steps}, and to correlate the user demand (traffic type and location) with the cells serving it. This analysis, that is seldom performed in the literature because of the amount and variety of input data it needs, allows us to draw some unexpected conclusions, summarized next.

\section{The role of LTE deployments}
\label{sec:findings}

\Fig{n-cells} summarizes the number of 3G and LTE cells that each operator has in Boston and Brooklyn. It is very interesting to observe that the number of cells and the fraction of LTE ones decidedly depends upon the operator. AT\&T and T-Mobile have a predominantly 3G network, with the latter deploying a larger number of cells than its competitors. Verizon, on the other hand, deploys mostly LTE base stations, partially because after the adoption of LTE they made substantial efforts to improve their coverage.

\subsection{LTE coverage}

\Fig{cov-boston} and \Fig{cov-brooklyn} show the distance between each 3G and LTE user (active or not) and the base station she is attached to. We can observe an unexpected and significant fact: LTE users tend to be much farther away from their base stations than 3G ones. This is consistent with the fact that LTE cells serve more traffic (\Fig{traffic}), but it contradicts the widespread belief that LTE is primarily used to improve network capacity.

Many studies on cellular network deployment are motivated by the expected increase in mobile data -- especially, but not only, video. Such increase can disrupt network connectivity, and therefore mobile operators need to deploy additional base stations (LTE, in this case) where the demand is higher. In these scenarios, operators would opt for high-capacity, low-coverage LTE cells, including small cells and femtocells. Instead, we observe a marked preference for large cells, suggesting that coverage is the operators' priority.

This is confirmed by \Fig{maps-brooklyn}, showing {\em where} each operator deploys LTE and 3G cells, as well as the coverage area thereof. Verizon's mostly-LTE deployment (\Fig{map-brooklyn-verizon}) is mostly based on large cells, a sign that the operator is aiming at covering the whole area with as few cells as possible. Similarly, Sprint (\Fig{map-brooklyn-sprint}) tend to place large LTE cells in those areas with few or no 3G cells of theirs -- again using LTE for the primary purpose of enhancing coverage.

Even AT\&T (\Fig{map-brooklyn-att}) and T-Mobile (\Fig{map-brooklyn-tmobile}), that deploy LTE cells in the those populated areas wherein their 3G coverage is already strong, show a clear preference towards large cells. In these cases, a major reason for providing LTE coverage is, so to say, providing LTE coverage, not to fall behind competition. In this case a primary motivation for deployment LTE is commercial rather than technical.

Moving to Boston in \Fig{maps-boston}, we can observe similar patterns. Operators have very different pre-existing 3G networks, but they all seem to prefer large LTE cells over small ones. AT\&T (\Fig{map-boston-att}) and Sprint (\Fig{map-boston-sprint}) clearly use LTE to complement 3G coverage, while T-Mobile (\Fig{map-boston-tmobile}) focuses on downtown areas. Verizon exhibits a stronger tendency to maximize its coverage through large LTE cells.

Interestingly, operators follow similar patterns in both cities. This suggests that their deployment decisions are the result of company-wide policies, as well as local conditions such as customer distribution or data demand.

It is important to stress that in both areas we study LTE deployment is {\em ongoing}. We have the unique opportunity to observe the {\em first} LTE base stations that operators deploy, and infer thence what aspect of LTE -- higher capacity, better coverage, etc -- they need most urgently. Furthermore, we also have the opportunity to compare LTE networks with a pre-existing 3G ones, assessing whether operators tend to replicate their previous-generation deployments, to complement them, or to follow a completely different strategy.

\subsection{The distance traveled LTE data}

The last question we seek to answer is whether, and to which extent, LTE networks are built for video traffic. We ascertain this by looking at how far away video users are from the base station {\em serving} them. Notice that, unlike in \Fig{coverage}, we take into account the actual traffic served by LTE base stations.

\Fig{serve-boston} and \Fig{serve-brooklyn} depict the distance that LTE video and non-video traffic travel from the user to the serving base station. To a shorter distance correspond a better quality and a higher throughput; therefore, if LTE networks were designed around a certain type of traffic, we would expect such traffic to travel a shorter distance.

However, no such clear pattern can be identified: sometimes video traffic travels a longer distance, e.g., with Verizon in \Fig{serve-boston}, other times there is no difference between the two. Also notice that sometimes the same operator has video traffic traveling a shorter distance than the rest in some cities and a longer one in others (e.g., Sprint). In summary, optimizing the service of video traffic seems not to be one of the main purposes of LTE deployments.

One compelling explanation is provided to us by \Fig{video}, depicting how video traffic is served -- overwhelmingly, through Wi-Fi. While it is true mobile video -- i.e., video consumed through smartphones and tablets -- is rapidly growing, our evidence suggests that only a tiny fraction thereof, about~92\% in Brooklyn and~90\% in Boston, is {\em cellular} video. Indeed, cable operators are already seeking remedies to this situation (sometimes controversial ones, such as the Netflix-Comcast deal~\cite{comcast}), while at this stage mobile operators seem to have other priorities.

\section{Conclusion and current work}
\label{sec:conclusion}

We argued that crowd-sourced datasets, obtained from users of smartphone applications, are a very good tool to understand how mobile networks are planned and used. We presented two fine specimens of this category, obtained from WeFi users, in \Sec{dataset}, and explained how we process such potentially overwhelming information in \Sec{process}.

We then set out to check some popular assumptions concerning the deployment of LTE networks against our data. Unexpectedly, we found in \Sec{findings} that improved network capacity is {\em not} the main reason why operators deploy LTE, and video is {\em not} the type of traffic they are designed to serve. Rather, operators seem to use LTE primarily to improve their coverage, deploying low-frequency, high-range cells at strategic locations, in both downtown and suburban areas.

At a more general level, our data suggest that traffic demand is but one of the factors shaping LTE deployments, and sometimes not even the main one. Understanding the deployment decisions made by mobile operators -- and foreseeing similar decisions for next-generation networks -- requires accounting such factors as pre-existing, previous-generation networks and commercial policies.

What we can observe through our traces are {\em early} deployments of LTE, which gives us the opportunity to grasp the priorities of operators, i.e., which of the multiple benefits of LTE they seek to obtain {\em first}. At the same time, this means that our results should be taken with a grain of salt: as an example, in the long term LTE networks will more likely replace 3G networks than complement them.

Current work includes more sophisticated analysis of the data demand, with the purpose of identifying positive and negative correlations between types of traffic. In parallel, we are are seeking to improve our outreach: while the datasets described in \Sec{dataset} cannot be released, owing to privacy concerns and non-disclosure agreements, we do plan to make an anonymized, aggregated version thereof available to the community.

\bibliographystyle{IEEEtran}
\bibliography{biblio}

\end{document}